\documentclass[conference]{IEEEtran}
\IEEEoverridecommandlockouts
\usepackage{cite}
\usepackage{amsmath,amssymb,amsfonts}
\usepackage{algorithmic}
\usepackage{graphicx}
\usepackage{textcomp}
\usepackage{xcolor}
\usepackage{subfig}
\usepackage{csquotes}
\def\BibTeX{{\rm B\kern-.05em{\sc i\kern-.025em b}\kern-.08em
    T\kern-.1667em\lower.7ex\hbox{E}\kern-.125emX}}
\begin{document}

\title{Enhanced document retrieval with topic embeddings}

\author{
\IEEEauthorblockN{Kavsar Huseynova}
\IEEEauthorblockA{\textit{Information Technology Department} \\
\textit{Baku Higher Oil School}\\
Baku, Azerbaijan \\
kavsar.huseynova.std@bhos.edu.az\\
0009-0007-0362-9591}
\and
\IEEEauthorblockN{Jafar Isbarov}
\IEEEauthorblockA{\textit{Department of Computer Science} \\
\textit{George Washington University}\\
Washington, D.C., the U.S. \\
jafar.isbarov@gwmail.gwu.edu\\
0000-0001-8404-2192 }
}
\maketitle

\begin{abstract}
Document retrieval systems have experienced a revitalized interest with the advent of retrieval-augmented generation (RAG). RAG architecture offers a lower hallucination rate than LLM-only applications. However, the accuracy of the retrieval mechanism is known to be a bottleneck in the efficiency of these applications. A particular case of subpar retrieval performance is observed in situations where multiple documents from several different but related topics are in the corpus. We have devised a new vectorization method that takes into account the topic information of the document. The paper introduces this new method for text vectorization and evaluates it in the context of RAG. Furthermore, we discuss the challenge of evaluating RAG systems, which pertains to the case at hand.
\end{abstract}

\begin{IEEEkeywords}
Information retrieval, document retrieval, text embeddings
\end{IEEEkeywords}

\section{Introduction}
\subsection{Retrieval-Augmented Generation}
Retrieval-augmented generation (RAG) systems allow us to create chatbots on large text corpora, such as law corpus, textbooks, and software documentation.
RAG was introduced by \cite{10.5555/3495724.3496517}. It has seen a sudden rise in popularity due to the availability of improved large language models (LLMs), especially since the release of LLaMA models \cite{touvron2023llama}.

RAG system works as follows: (1) Text corpus is split into chunks and each chunk is vectorized. (2) User query is vectorized. (3) A similarity search is performed to find the chunk closest to the vectorized query. (4) The retrieved chunk is fed to the LLM along with the user query. (5) LLM uses this input to generate a free-form response.

One of the main bottlenecks in the performance of RAG systems is the retrieval step. The accuracy of the similarity search depends on multiple factors, including the choice of the similarity search algorithm, embedding method, and size of the indexed corpus. Corpus size becomes especially problematic if we are dealing with very similar documents. 

\subsection{Proposal}
In most cases, we are not dealing with raw text data. Instead, the text comes with ample metadata that can be used to boost the performance of the similarity search. Our proposal is to use the available topic information of each chunk (i.e. document) and add it to the retrieval process. We have successfully implemented two such methods in industrial settings, and this paper attempts to generalize and evaluate these methods. The first method relies on creating a new document embedding by combining the original document embedding with the topic embeddings generated from the entire topic. The second method consists of two steps (1) find the topic, and (2) find the document within the topic.

Our contributions in this paper are as follows:
\begin{enumerate}
    \item Propose two new methods for using topic metadata during the document retrieval process.
    \item Evaluate the suggested methods with calculating the distance measures on embedded documents.
    \item Propose a detailed problem statement for the next stages of this work.
\end{enumerate}

The paper is structured as follows: The next section contains a review of relevant literature on these topics. The third section provides a detailed explanation of the proposed methods. The fourth section consists of two parts. First part outlines our experiments, their explanations, and results. A comprehensive discussion of our results, the main shortcomings of our work, as well as suggestions for future research take place in the second part. We conclude with the paper in the fifth section.

\section{Related Work}

\begin{figure*}    
    \centering
    \includegraphics[width=1.5\columnwidth]{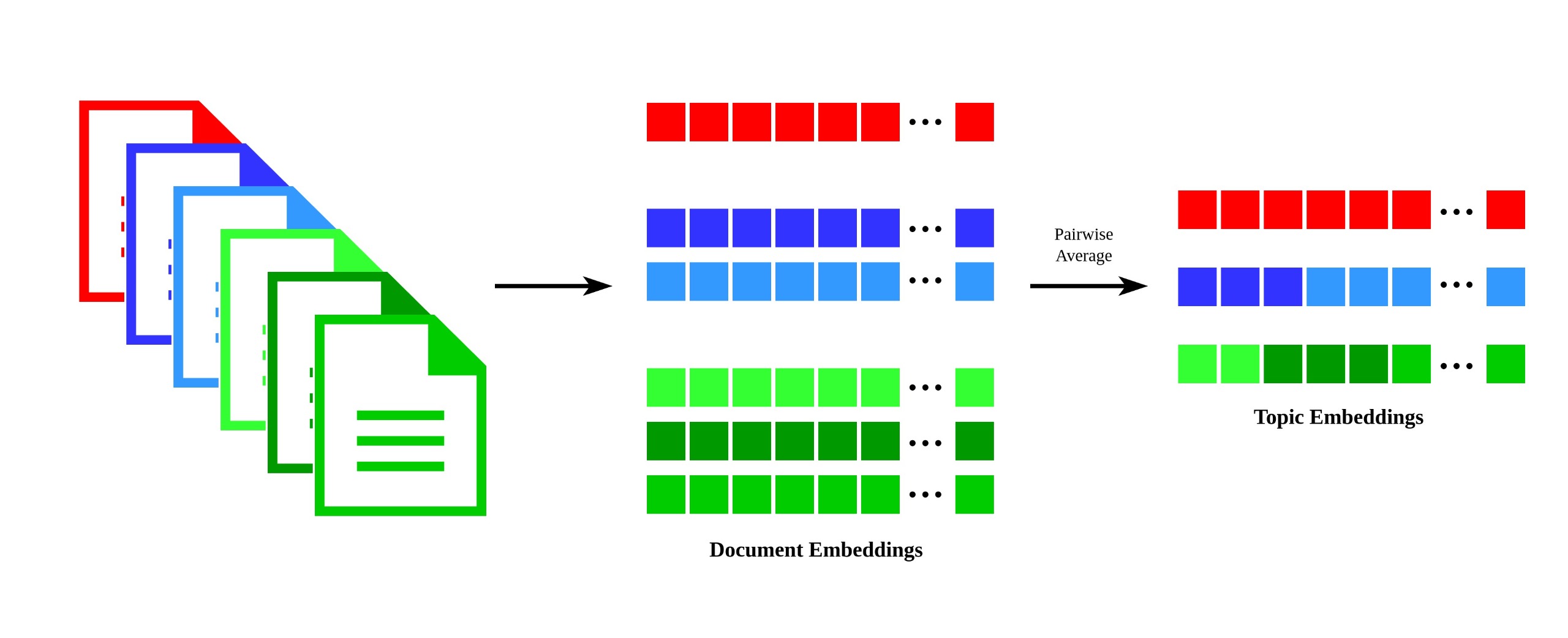}
    \caption{Process for generating topic embeddings from original documents.}
    \label{fig:topic}
\end{figure*}

RAG was introduced by \cite{10.5555/3495724.3496517} in 2020. Their work underpins more advanced RAG systems that we see today. \cite{Liu2020RetrievalAugmentedGF} uses this idea to create a code summarization tool. Multimodal RAG systems are also seeing increasing popularity. \cite{10.5555/3618408.3620067} implements a competitive image and caption generation system with RAG architecture.

Vectorization is an integral part of RAG systems. Traditional vectorization techniques include bag-of-words \cite{8950616}, which simply uses word frequency and term frequency-inverse document frequency \cite{Qaiser2018TextMU}, which additionally takes into account how unique the word is to a particular document.

Relatively modern approaches include Word2Vec, GloVe, and fastText. Word2Vec \cite{Mikolov2013EfficientEO} can use either continuous bag-of-words or continuous skipgram models. GloVe \cite{pennington-etal-2014-glove} is a log-bilinear regression model that was trained on a word-word cooccurrence matrix. fastText \cite{Bojanowski2016EnrichingWV} is a skip-gram model that treats words as a combination of character n-grams.

The latest family of text vectorization methods relies on transformer-based language models. BERT \cite{Devlin2019BERTPO} is a textbook example of this. These models usually use a masked attention mechanism to learn the context of tokens in a text. This information can later be used in various tasks, such as named entity recognition and sentiment analysis. The main advantage of this approach is that embeddings are contextual, i.e. they change depending on their place in the text. This is not the case with the aforementioned methods. 

\cite{NIPS2017_59dfa2df} exploits inherent hierarchical structure in the data during the embedding process. Our work is similar but distinct: We use explicit hierarchy, not implicit. As far as we know, no work has attempted this. \cite{8658633} attempts a similar approach to enhance the performance of image classification models on ImageNet.

All of the mentioned works are either language-agnostic, or English-specific. Our dataset is in Azerbaijani, therefore works for the Azerbaijani language are of special interest to us.

Among the open-source embedding models, some claim to have an understanding of the Azerbaijani language. Google has released a multilingual version of the famous BERT model. We are aware of several unpublished attempts to use this model (directly or by fine-tuning) for various NLP tasks in Azerbaijani, all of them with limited success. Due to this, we avoided using the model as an embedder. Another multilingual model that has some understanding of Azerbaijani is mGPT. This is an unofficial version of the GPT-2 model released by \cite{shliazhko2023mgpt} that was trained in a text corpus consisting of multiple languages. Its text generation capabilities in Azerbaijani are not satisfactory. This is why we did not use it in the generation part of our RAG systems.

\section{Methods}

We propose two new methods that use topic information during document retrieval: topic-enhanced document embeddings, and two-stage document retrieval.

\subsection{Topic-enhanced document embeddings}
Here, we create topic embeddings and then use these embeddings to update the original document embeddings in one of two ways. 

\textbf{Step 1: Create document embeddings.}

This is part of the traditional RAG pipeline. We simply chunk the document and embed these chunks separately.

\textbf{Step 2: Separate documents into topics.}

We assume the topic information is provided either implicitly or explicitly. In any case, we need to have explicit topic labels by the end of this stage.

\textbf{Step 3: Creating topic embeddings.}

Topic embeddings are supposed to be a vector of the same size as document embeddings. If we are using a neural network like BERT to embed the text, we usually cannot feed the entire topic into the model. We can bypass this by taking an element-wise average of all document embeddings for that topic. If we are using a statistical method like TF-IDF, we can run it on the entire text regardless of the text length. However, we expect both the original document embeddings and the topic embeddings to use the same embedding method. Figure \ref{fig:topic} visualizes the process of obtaining topic embeddings from documents.

\textbf{Step 4: Update the document embeddings}

We propose two alternate versions here:

\begin{enumerate}
    \item \textit{Average method}: Take an average of document embeddings and topic embeddings.
    \item \textit{Append method}: Concatenate document embeddings and topic embeddings. 
\end{enumerate}
The second method creates a new problem because now the embedding dimension of a query does not match the dimensions of our embedded documents. To solve this problem, we can duplicate the length of our query embeddings. 

\subsection{Two-stage document retrieval}
This is a much simpler approach. We create topic embeddings just like the first method, but we perform retrieval in two stages:
\begin{enumerate}
    \item Retrieve a topic based on the topic-embeddings.
    \item Retrieve a document within that topic.
\end{enumerate}
However, two-stage retrieval system comes with own challenges. One of which is increased inference time. 

\section{Results}

\subsection{Experiments}

To evaluate these approaches, we have curated a dataset from Azerbaijani laws. We have selected 14 major laws and split them into chunks of 2000 characters. You can find these laws in Table \ref{laws}. OpenAI's "text-embedding-3-small" embedding model has been used to embed the chunks. Each law is treated as a distinct topic. We then used both the average method and the append method to update the document embeddings.

\begin{table}[]
\centering
\caption{Azerbaijani laws in our dataset.}
\begin{tabular}{|l|l|}
\hline
\textbf{Topic} & \textbf{Chunk Count} \\ \hline
Criminal Code & 716 \\ \hline
Code of Criminal Procedure & 1035 \\ \hline
Customs Code & 280 \\ \hline
Constitution & 100 \\ \hline
Forest Code & 60 \\ \hline
Civil Procedure Code & 551 \\ \hline
Civil Code & 1283 \\ \hline
Migration Code & 181 \\ \hline
Water Code & 76 \\ \hline
Land Code & 150 \\ \hline
Tax Code & 1122 \\ \hline
Code on Administrative Violations & 781 \\ \hline
Labor Code & 414 \\ \hline
Education Law & 129 \\ \hline
\end{tabular}
\label{laws}
\end{table}

By treating topics (i.e., laws) as cluster labels, we have evaluated the "clustering results" of three methods:
\begin{itemize}
    \item original embeddings
    \item averaged embeddings
    \item appended embeddings
\end{itemize}

\begin{figure*}    
    \centering
    \subfloat[]{\label{normal}\includegraphics[width=\columnwidth]{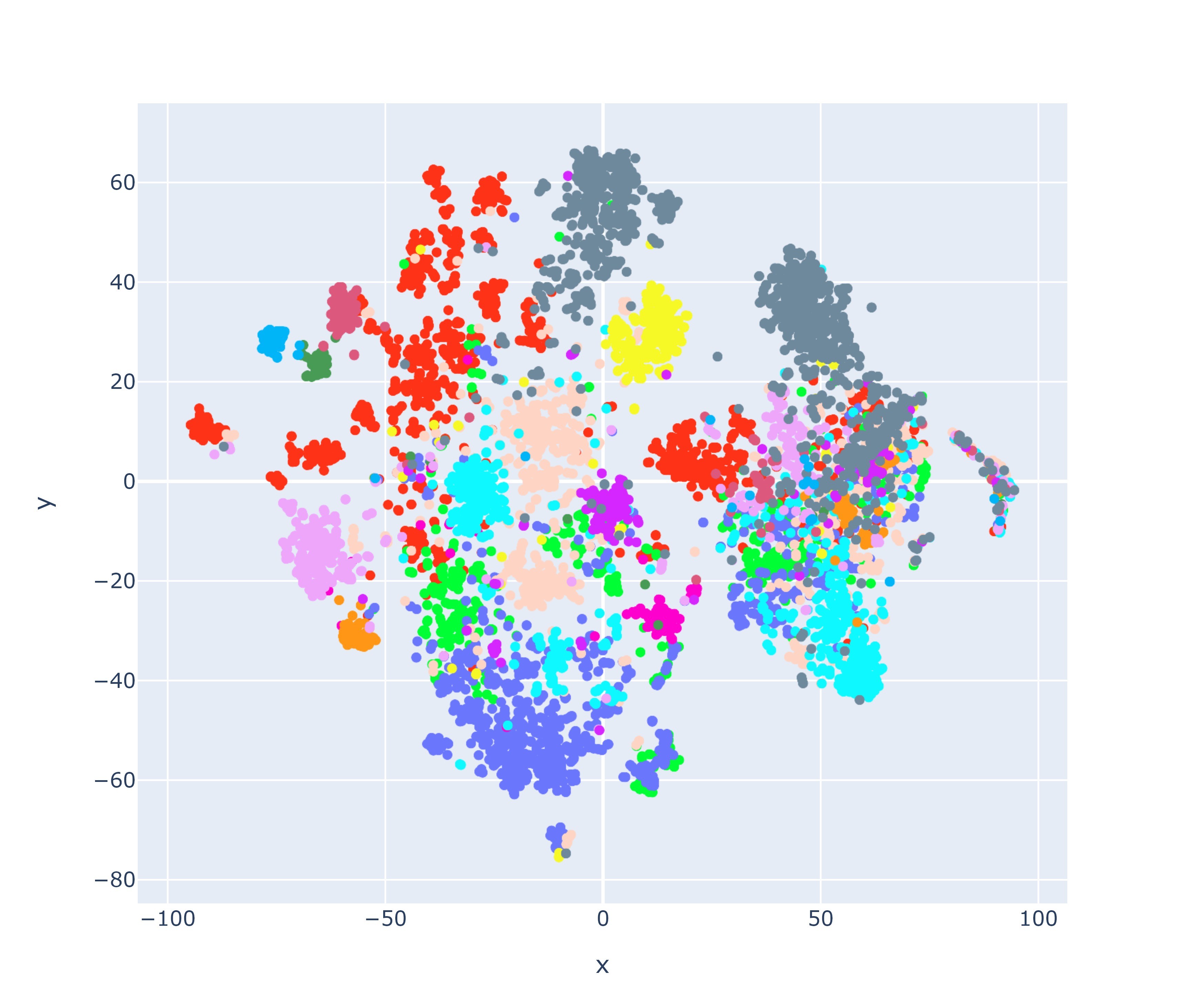}} 
    \subfloat[]{\label{avg}\includegraphics[width=\columnwidth]{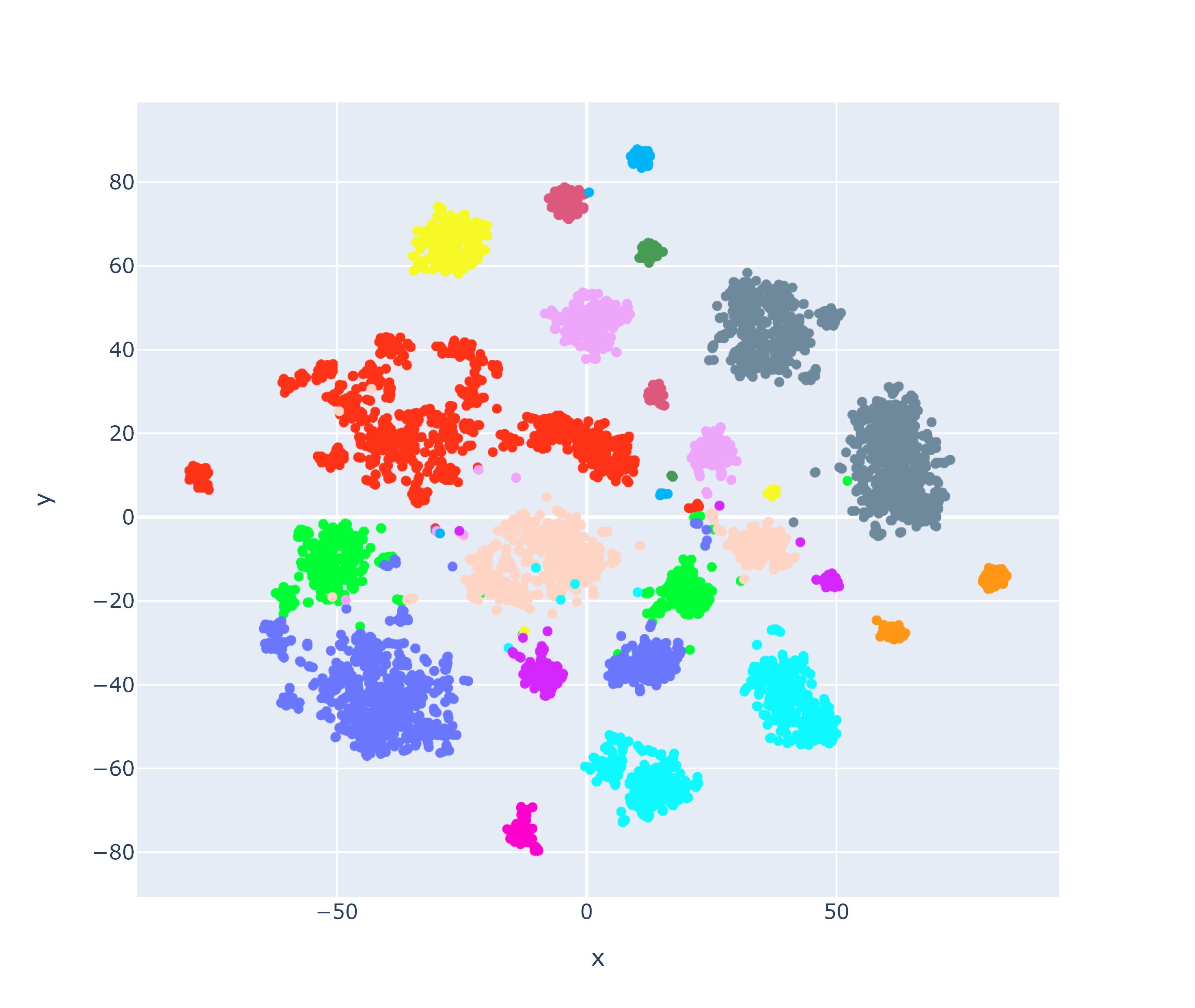}} \\
    \subfloat[]{\label{append}\includegraphics[width=\columnwidth]{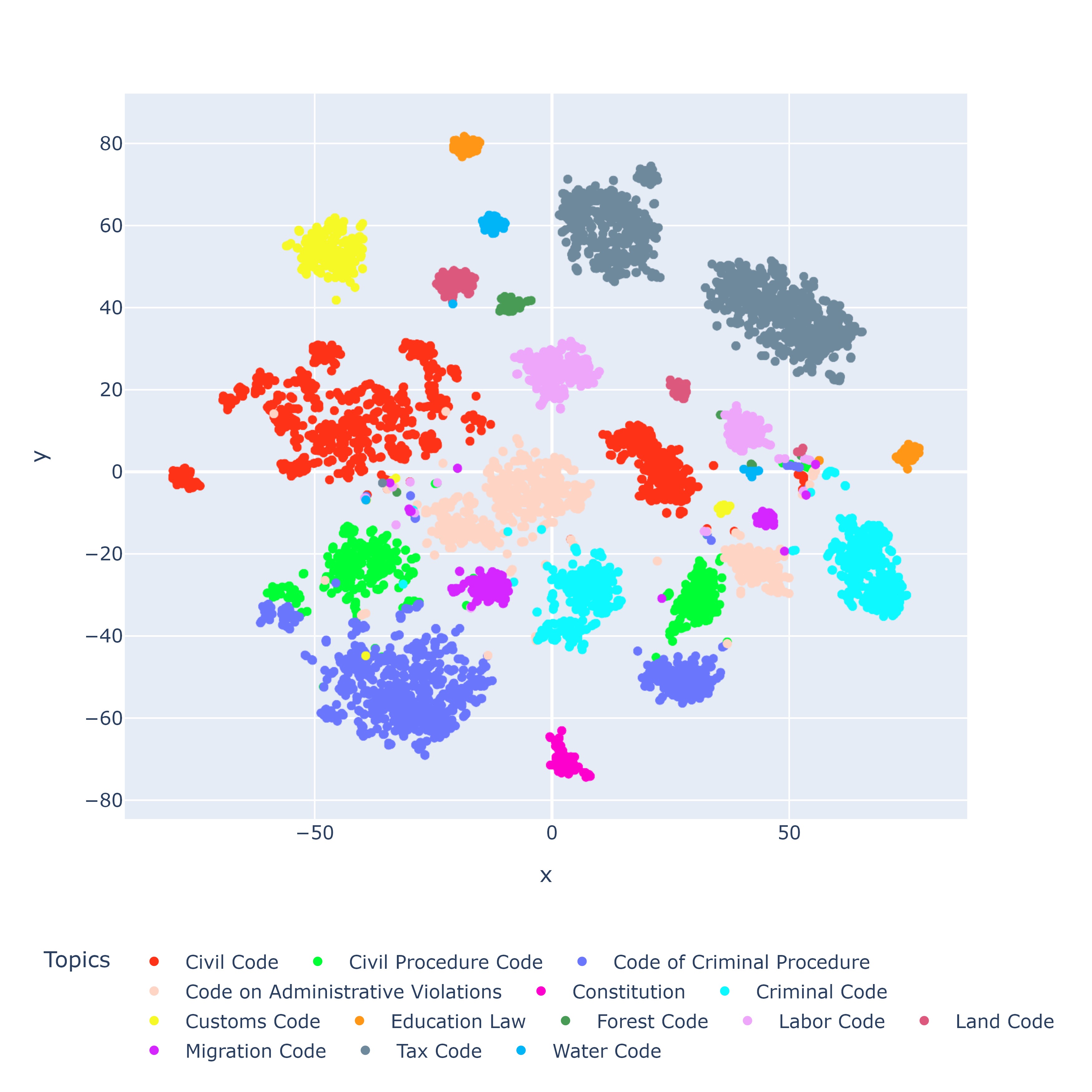}}
    \caption{2D visualization of topics with (a) original embeddings, (b) averaged embeddings, (c) appended embeddings.}
    \label{globallable}
\end{figure*}

To visualize the embeddings of different methods, the tSNE algorithm has been used to reduce the number of dimensions to 2 \cite{10.5555/3586589.3586890}. You can see the results in Figure \ref{normal} and \ref{avg}.

To accurately evaluate our proposed methods, it is essential to use a dedicated RAG evaluation dataset. Although we initially attempted to construct this dataset synthetically, we recognized the limitations of this approach and concluded that a natural evaluation dataset is necessary. We have explored this issue in detail in the section B.  

Another approach to calculate the performance of these "clustering models" is using the cluster validity indices. These indices provide insights into how well the embeddings separate different topics, which is crucial for ensuring that similar documents are grouped effectively—a key factor in improving retrieval performance.

We have used three different indices:
\begin{itemize}
    \item Silhouette Coefficient
    \item Davies–Bouldin index
    \item Calinski–Harabasz index
\end{itemize}

The Silhouette Coefficient, ranging from -1 to +1, measures clustering effectiveness, with higher values indicating better clustering. In contrast, the Davies-Bouldin Index (DBI) assesses cluster separation and compactness, where lower values signify better quality. The Calinski-Harabasz (CH) Index evaluates how well-separated and dense clusters are, with higher values indicating better clusters, typically identified by a peak in the index. Together, these metrics provide a comprehensive assessment of clustering performance.

The results are available in Table \ref{clustering_metrics}. As you can see, adding topic embeddings to the original embeddings results in better separation of different topics. The average method performs better than the append method, although we suspect that it may be data-specific. As future work, these experiments can be performed on larger and more variable datasets to assess the methods' performance against each other.

We were unable to evaluate the 'Two-stage document retrieval' method due to limitations in the evaluation dataset, which will be discussed further.

\begin{table}
\centering
\caption{Performance of original and topic-based clustering. DBI: Davies–Bouldin index, CHI: Calinski–Harabasz index.}
\label{clustering_metrics}
\begin{tabular}{|l|l|l|l|}
\hline
Metric                  & Original & Average & Append \\ \hline
Silhouette        &   0.01   & \textbf{0.11} & 0.06 \\ \hline
DBI    &   4.60   & \textbf{2.30} & 3.25 \\ \hline
CHI &  63.42  & \textbf{253.67} & 126.84 \\ \hline
\end{tabular}
\end{table}

\subsection{Challenges and limitations}

The main challenge we are facing is the lack of an end-to-end evaluation dataset. This dataset needs to have the following features:
\begin{itemize}
    \item A text corpus with topic labels
    \item Natural queries that can be answered based on a part of this corpus.
\end{itemize}
Using this dataset, one can first embed the queries with traditional and our proposed methods, then run the similarity search on each of them to match the queries with respective documents, and hence get an evaluation score to make a fair comparison between traditional RAG system and our approach.

This first feature is easy to attain. We have already performed some analysis on such datasets. We have attempted to gain the second feature by synthetic query generation, but it did not achieve any better result than normal RAG system, because generated queries are highly specific to the chunk provided. For example, we used the following query:
\begin{displayquote}

\begin{it}
"\textbf{Given the following context, generate a question that would be asked by a curious citizen:}

Context:

Article 25. Right to equality

I. Everyone shall be equal before the law and the courts.

II. Men and women possess equal rights and freedoms.

III. The State shall guarantee the equality of rights and freedoms to everyone, irrespective of race, ethnicity, religion, language, sex, origin, property status, occupation, beliefs or affiliation with political parties, trade union organisations or other public associations. Restrictions of rights and freedoms on the grounds of race, ethnicity, religion, language, sex, origin, beliefs, or political or social affiliation are prohibited.

IV. No one may be harmed, granted advantages or privileges, or refused to be granted advantages and privileges on the grounds laid down in Paragraph III of the present Article.

V. Everyone shall be guaranteed equal rights in any proceeding before state authorities and bearers of public authority that decide upon his/her rights and duties.

VI. Persons with impaired health are entitled to all rights and carry all duties enshrined in this Constitution, except in cases when enjoyment of rights and performance of duties is impeded by their limited abilities."
\end{it}

\end{displayquote}

This resulted in the following response:
\begin{displayquote}
\begin{it}
    "What measures are in place to ensure enforcement of Article 25, particularly regarding equality before the law and in courts, as well as the guarantee of equal rights and freedoms regardless of various personal characteristics?"
\end{it}
\end{displayquote}

As you can see, if we generate a large dataset automatically using these prompts, we will end up with questions that exactly match the provided context (i.e., chunk). Because the content of the queries are highly related with the document chunks, our proposed methods achieve similar performance as the traditional RAG system.

Although the synthetic query generation attempts did not work, we have tried it on the English version of our dataset. As far as we are concerned, no LLM can generate natural text in Azerbaijani at a level that would be sufficient for this task. Due to these restrictions, we are researching the possibility of creating a natural dataset. We can use logs of one of our chatbots in production for this, or we can create an annotator team to build the dataset.

Another inherent limitation of our approach is that it depends on the existence of topic metadata. It would be interesting to devise a method to infer the topic information from the raw data itself, although we do not believe that there is a generalizable approach to this problem.

\section{Conclusion}
This paper introduces an enhanced vectorization technique for document retrieval. As RAG applications become increasingly popular, it is important to handle the challenge of having a large number of documents within the same database. We proposed two new methods that utilize the provided topic information in the text corpus. We have shown that by implementing our two novel approaches, the distance between different documents can be broadened through the introduction of topic embeddings, potentially resulting in more accurate similarity searches. Despite our progress in RAG application, there is a need to refine evaluation techniques. To ensure accurate and representative performance assessments of our research, an end-to-end evaluation dataset should be curated. As future work, evaluating our method in multiple languages would better demonstrate its generality.

\section*{Acknowledgements}

The authors declare that this manuscript is original, has not been published before, and is not currently being considered for publication elsewhere.

\bibliographystyle{IEEEtran}
\bibliography{refs}

\vspace{12pt}
\end{document}